\documentclass{aa}
\usepackage{graphicx}
\usepackage{txfonts}
\usepackage{natbib}
\usepackage{color}
\usepackage[colorlinks, allcolors=blue]{hyperref}

\begin{document}

\title{The formation of small-scale umbral brightenings\\ in sunspot atmospheres}

\author{C. J. Nelson$^{1,2}$, V. M. J. Henriques$^{1,3}$, M. Mathioudakis$^1$, F. P. Keenan$^1$}

\offprints{c.j.nelson@sheffield.ac.uk}
\institute{$^1$Astrophysics Research Centre (ARC), School of Mathematics and Physics, Queen’s University, Belfast, BT7 1NN,\\ Northern Ireland, UK.\\
$^2$Solar Physics and Space Plasma Research Centre (SP2RC), University of Sheffield, Hicks Building, Hounsfield Road,\\ Sheffield, S3 7RH, UK.\\
$^3$Institute of Theoretical Astrophysics, University of Oslo, PO Box 1029 Blindern, 0315 Oslo, Norway.}

\date{}

\abstract
{Sunspot atmospheres have been shown to be highly inhomogeneous hosting both quasi-stable and transient features, such as small-scale umbral brightenings (previously named `umbral micro-jets') and dark fibril-like events.}
{We seek to understand the morphological properties and formation mechanisms of small-scale umbral brightenings (analogous to umbral micro-jets). In addition, we aim to understand whether links between these events and short dynamic fibrils, umbral flashes, and umbral dots can be established.}
{A Swedish $1$-m Solar Telescope (SST) filtergram time-series sampling the \ion{Ca}{II} H line and a {\it CRisp Imaging Spectro-Polarimeter} (SST/CRISP) full-Stokes $15$-point \ion{Ca}{II} $8542$ \AA\ line scan dataset were used. The spatial resolutions of these datasets are close to $0.1$\arcsec\ and $0.18$\arcsec\ with cadences of $1.4$ seconds and $29$ seconds, respectively. These data allowed us to construct light-curves, plot line profiles, and to perform a weak-field approximation in order to infer the magnetic field strength.}
{The average lifetime and lengths of the $54$ small-scale brightenings identified in the sunspot umbra are found to be $44.2$ seconds ($\sigma$=$20$ seconds) and $0.56$\arcsec\ ($\sigma$=$0.14$\arcsec), respectively. The spatial positioning and morphological evolution of these events in \ion{Ca}{II} H filtergrams was investigated finding no evidence of parabolic or ballistic profiles nor a preference for co-spatial formation with umbral flashes. Line scans in \ion{Ca}{II} $8542$ \AA\ and the presence of Stokes V profile reversals provided evidence that these events could form in a similar manner to umbral flashes in the chromosphere (i.e. through the formation of shocks either due to the steepening of localised wavefronts or due to the impact of returning material from short dynamic fibrils, a scenario we find evidence for). The application of the weak-field approximation indicated that changes in the line-of-sight magnetic field were not responsible for the modifications to the line profile and suggested that thermodynamic effects are, in fact, the actual cause of the increased emission. Finally, a sub-set of small-scale brightenings were observed to form at the foot-points of short dynamic fibrils.}
{The small-scale umbral brightenings studied here do not appear to be jet-like in nature. Instead they appear to be evidence of shock formation in the lower solar atmosphere. We found no correlation between the spatial locations where these events were observed and the occurrence of umbral dots and umbral flashes. These events have lifetimes and spectral signatures comparable to umbral flashes and are located at the footpoints of short dynamic fibrils, during or at the end of the red-shifted stage. It is possible that these features form due to the shocking of fibrilar material in the lower atmosphere upon its return under gravity.}

\authorrunning{C. J. Nelson et al.}
\titlerunning{Structuring In Umbral Chromospheres}

\maketitle

\section{Introduction}

Spatial inhomogeneities on a variety of scales have been documented within sunspot atmospheres despite the apparently near-uniform nature of the local magnetic field (\citealt{Keppens96,Solanki03}). Relatively large-scale phenomena such as umbral flashes (see, for example: \citealt{Beckers69,Wittmann69,vanderVoort03}) and running penumbral waves ({\it e.g}, \citealt{Zirin72,Giovanelli72,Freij14}) have been observed for decades, and are widely attributed to the steepening of acoustic or magneto-acoustic waves as they propagate upwards into the atmosphere along the natural guide provided by the magnetic field of the sunspot itself (\citealt{Bard10}). Numerous other smaller-scale features have also recently been discovered, such as short dynamic fibrils (\citealt{vanderVoort13}), chromospheric spikes (\citealt{Yurchyshyn14}) and umbral micro-jets (\citealt{Bharti13}). These features indicate the presence of a range of complex physical processes, such as shocks and (potentially) magnetic reconnection, occurring on sub-arcsecond scales making sunspot chromospheres some of the most interesting solar regions to study.

\begin{figure*}
\includegraphics[scale=0.48]{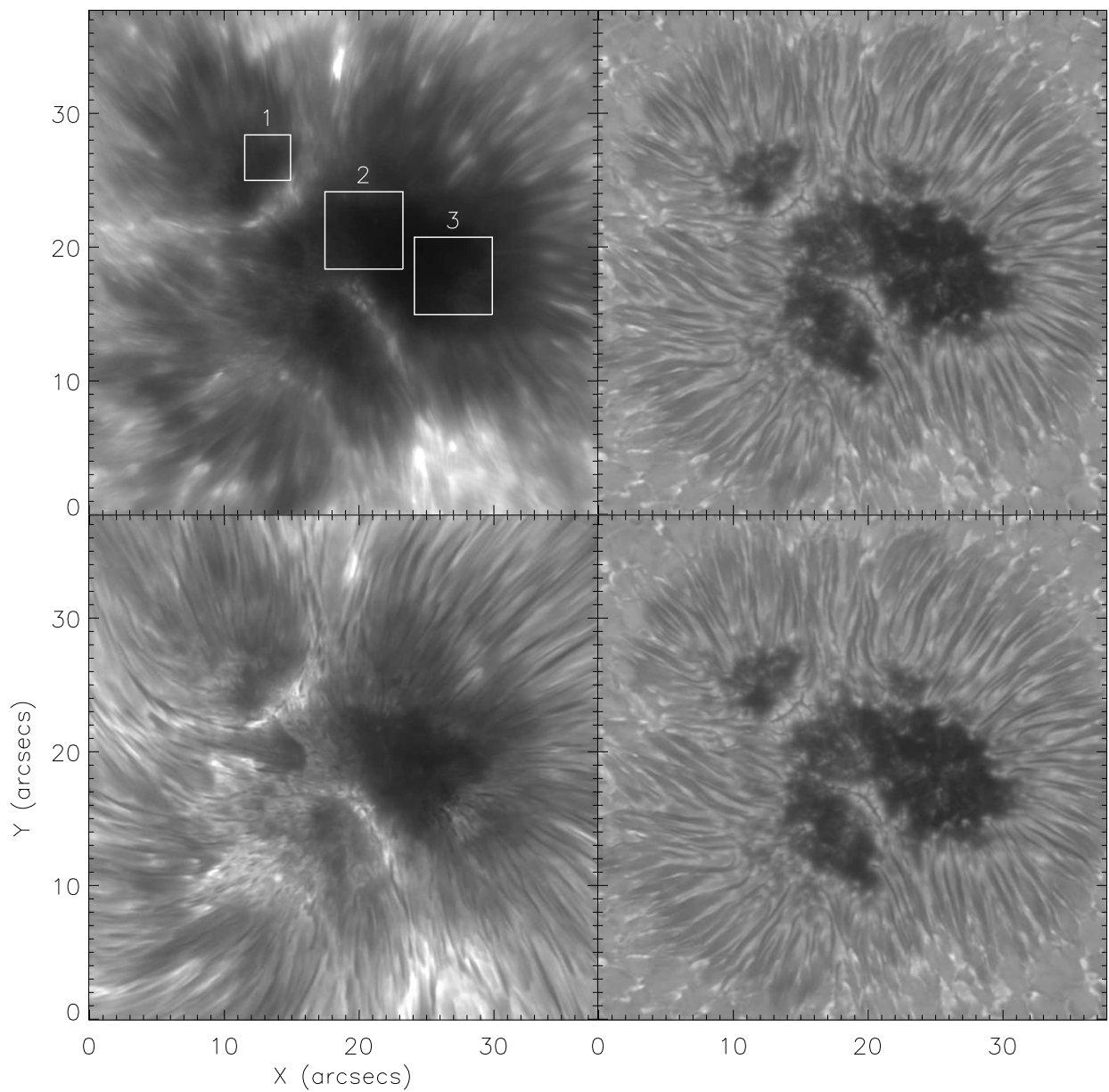}
\caption{Context image of the sunspot analysed in this investigation at $10$:$49$:$52$ UT. Plotted clockwise from the top left are: a \ion{Ca}{II} H line core filtergram image, the blue wing ($0.942$ \AA) of the \ion{Ca}{II} $8542$ \AA\ line; the red wing ($-0.942$ \AA) of the \ion{Ca}{II} $8542$ \AA\ line; and the \ion{Ca}{II} $8542$ \AA\ line core. The numbered white boxes in the top left frame indicate the regions of the umbra analysed in detail in  this paper.}
\label{Context}
\end{figure*}

Thin dark umbral fibrils with large horizontal extents were first observed by \citet{Socas09} who studied regions highlighted by umbral flashes in \ion{Ca}{II} H line core filtergrams. Evidence of fine-scale structuring had previously been inferred through spectro-polarimetric measurements by \citet{Socas00b}, \citet{Socas00a}, and \citet{Centeno05}. These authors discussed the contradiction between the existence of these events and the concept of predominantly vertical magnetic fields within the umbra itself. \citet{Henriques13} corroborated the presence of these features and found dark fibrils with similar properties over the unflashed penumbra, indicating that the umbral fibrils could be similar in nature to the penumbral ones and thus stable rather than transient. It was hypothesised that these events were formed in the upper photosphere and that they were not vertical in nature. In an extension of this work, \citet{Henriques15} identified a sample of these features in three different sunspots and confirmed their stability over the course of at least two umbral flashes. In that work the largest feature extended up to 3750 km and showed changes in orientation throughout its body. They found a partial match between small-scale umbral fibrils and features in H$\alpha$, meaning the shorter umbral fibrils those authors discussed could possibly be the short dynamic fibrils identified by \citealt{vanderVoort13}. 

\begin{figure*}
\includegraphics[scale=0.55,trim={1cm 0 1cm 0}]{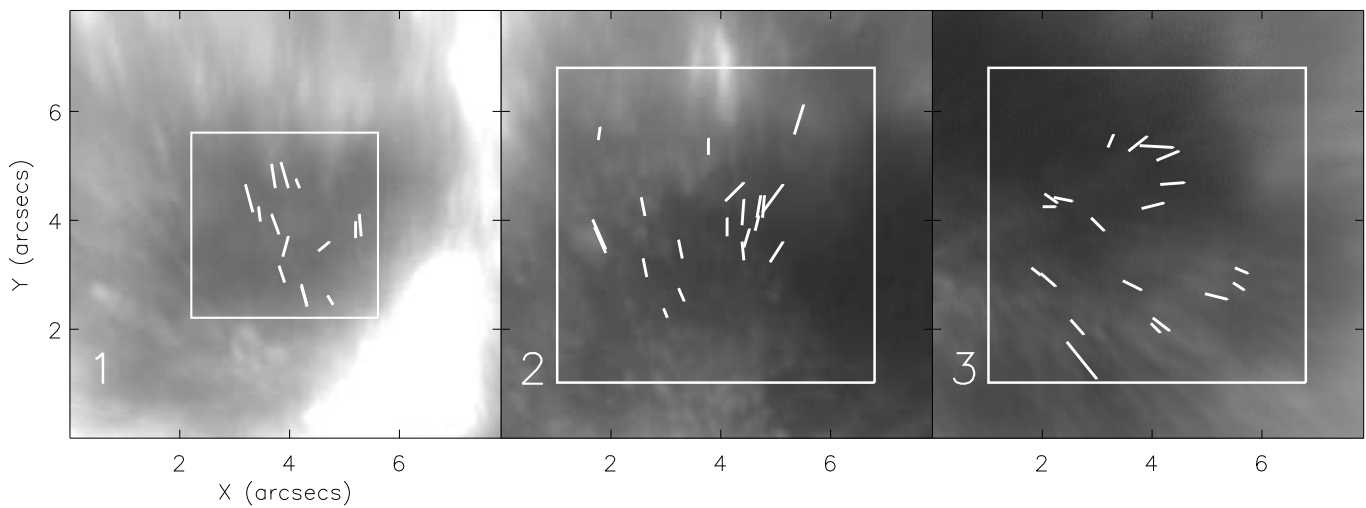}
\caption{White lines indicate the spatial positioning of all $54$ SSUBs identified within these \ion{Ca}{II} H data at their peak length. The white boxes indicate the sub-regions plotted in Fig.~\ref{Context} in which the selection of these features was attempted. The background displays each FOV at the initial time-step.}
\label{Position}
\end{figure*}

Short dynamic fibrils, or sunspot fibrils, were first identified by \citet{vanderVoort13} who observed relatively short features whose tips followed parabolic trajectories with maximum extensions of around $1000$ km. The distribution of the properties of these events followed a similar pattern to active region short dynamic fibrils (see, for example, \citealt{Hansteen06,dePontieu07}), although the lengths and lifetimes of the sunspot events were, in general, smaller. It was suggested that the parabolic evolution profiles of these events, observed in both \ion{Ca}{II} $8542$ \AA\ and H$\alpha$ data, could be linked to the corrugated propagation of waves or shocks upwards through the solar atmosphere leading to the formation of jets (see, for example, \citealt{Heggland07,Heggland11}). This mechanism was also suggested to be the cause of the chromospheric spikes visible in H$\alpha$ observations discussed by \citet{Yurchyshyn14}. The properties of these chromospheric spike events were very similar to those discussed by \citet{vanderVoort13} with lengths close to $1000$ km and widths of around $100$ km. Thus, they are likely the same phenomenon.

A more transient phenomenon was identified in \ion{Ca}{II} H filtergrams by \citet{Bharti13}, who named their new class of feature the `umbral micro-jet'. These events were described as small-scale (extents below one arc-second), short-lived (lifetimes of less one minute) brightenings against the dark umbra, and were hypothesised to be the umbral counterpart of the penumbral micro-jets widely associated with magnetic reconnection (e.g. \citealt{Katsukawa07,Jurcak08,Vissers15}). However, the $19$ second cadence data analysed by \citet{Bharti13} only sampled each individual micro-jet around two or three times.  This meant their full dynamical behaviour (and e.g. confirmation of a jet-like morphology), which could offer valuable clues as to the formation mechanism responsible for these events, could not be inferred. Furthermore, both \citet{vanderVoort13} and \citet{Yurchyshyn14} predicted that these umbral microjets were dissimilar to the short dynamic fibrils and spikes identified in their respective researches, although neither analysed co-temporal \ion{Ca}{II} H data. Higher cadence data and more comprehensive spectral sampling (e.g. co-spatial \ion{Ca}{II} H filtergrams and \ion{Ca}{II} $8542$ \AA\ line scans) would provide one route for identifying any potential relationships between these events, and further understanding the complex physics occurring within sunspot atmospheres.

In this article, we identify a sample of small-scale umbral brightenings (SSUBs) in a \ion{Ca}{II} H line core time-series and measure their properties. We note that these events are analogous to the umbral micro-jets discussed by \citet{Bharti13} but named using a different convention to avoid the jet terminology. This work aims to test the hypothesis that these events are reconnection driven in addition to inferring any links between these events and small-scale umbral flash structuring, umbral dots, and short dynamic fibrils. In Section~\ref{Observations}, we present the observations, data reduction techniques used, and our feature selection methodology. Section~\ref{Results} introduces the results obtained through analysis of these SSUBs, including basic statistics and morphological properties. Spectro-polarimetric signatures and evidence of sunspot fibrils co-spatial to these events are also included. Section~\ref{Conclusions} contains our conclusions and a discussion of how these events fit into our understanding of dynamic sunspot umbrae.

\section{Observations}
\label{Observations}

The data analysed here were obtained using the Swedish $1$-m Solar Telescope (SST; \citealt{Scharmer03}) on $28$th July $2014$ between $10$:$43$:$44$ UT and $11$:$24$:$34$ UT. A large sunspot within AR $12121$ (situated at initial coordinates of $x_\mathrm{c}$=$76.5$\arcsec, $y_\mathrm{c}$=$46.5$\arcsec\ with respect to the disc centre) was selected for observing using the standard set-up, in which the light-beam was split into blue and red components. One camera, with a passband of $1.1$ \AA\ FWHM centred on the \ion{Ca}{II} H line-core ($3968.4$ \AA), was placed in the blue beam. Data obtained with this camera were reconstructed using the MOMFBD technique (\citealt{vanNoort05}), which returned a time-series with a final science-ready cadence of $1.4$ seconds and pixel scale of approximately $0.034$\arcsec. This time-series contained $1797$ frames of generally high quality data. For a large number of frames, the reconstructed resolution approached the diffraction limit of $0.1$\arcsec. 

The red beam was sampled using the {\it CRisp Imaging Spectro-Polarimeter} (SST/CRISP; \citealt{Scharmer06,Scharmer08}) instrument, which ran a sequence involving a $15$-point full-Stokes scan sampling unevenly between $\pm0.942$ \AA\ of the \ion{Ca}{II} $8542$ \AA\ line at positions of [$\pm0.942$ \AA, $\pm0.580$ \AA, $\pm0.398$ \AA, $\pm0.290$ \AA, $\pm0.217$ \AA, $\pm0.145$ \AA, $\pm0.073$ \AA, and $0$ \AA] (with respect to the line core), as well as imaging of the H$\alpha$ line core. In addition, wide-band images were collected for each scan to provide photospheric context. The data were reduced using the CRISPRED package (\citealt{delaCruzRodriguez15}), including the methods discussed by \citet{Henriques12}, in order to minimise residual seeing impact on the profiles, and analysed, in part, using the CRISPEX tool (\citealt{Vissers12}). The cadence of this routine, which was repeated $79$ times in total, was approximately $29$ seconds, and the pixel scale of these red beam images was approximately $0.059$\arcsec. As with the blue beam, the resolution of these data often appeared to approach the diffraction limit of $0.18$\arcsec.

\begin{figure*}
\includegraphics[scale=0.47,trim={1.7cm 0 1.7cm 0}]{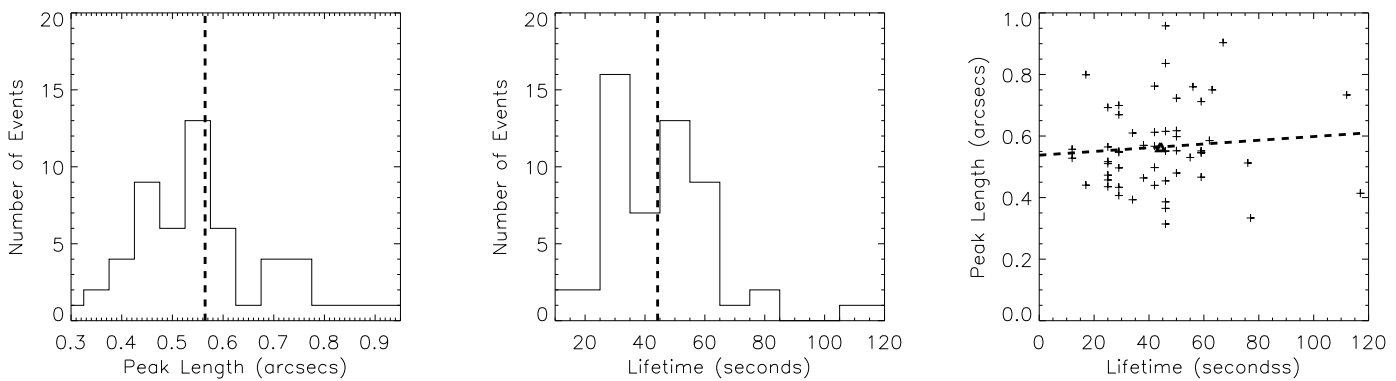}
\caption{Basic properties of the sample of SSUBs, inferred from \ion{Ca}{II} H data. The left and centre panels plot histograms of peak length and lifetime, respectively, with the dashed lines in each frame corresponding to the appropriate mean values. The right panel displays a scatter plot of peak length against lifetime, with the dashed line indicating a least squares regression linear fit calculated for these data (with a rate of change of an extra $0.0006$\arcsec\ of length per extra second of lifetime). The triangle plots the position of the mean of both variables.}
\label{Stats}
\end{figure*}

A region of the entire field-of-view (FOV), sampled at approximately $10$:$49$:$52$ UT, is plotted for reference in Fig.~\ref{Context}. The top left frame depicts a \ion{Ca}{II} H filtergram image and includes three numbered boxes which outline the regions of the umbra selected for in-depth analysis in this investigation. Other panels, clockwise from top right, plot the \ion{Ca}{II} $8542$ \AA\ blue ($0.942$ \AA) wing, red ($-0.942$ \AA) wing, and line core. The blue and red wing images both highlight the non-uniformity of the sunspot with a complete light-bridge separating a small portion of umbra in the top left of the FOV, as well as another incomplete excursion into the main umbra apparent in the centre of the FOV. A range of dynamic features are easily identified within movies of this time-series (including umbral flashes). However, the overall shape of the sunspot remains relatively stable throughout the analysed $40$ minute period.

The small-scale, short-lived nature of the events analysed here, combined with the dynamic background of the sunspot chromosphere, presents numerous complexities which hinder unambiguous feature detection. Initially, three sub-regions of the umbra (indicated by a white box in Fig.~\ref{Context}) were selected before an array was created containing the data recorded at these locations at every tenth time-step (mirroring the cadence of the data analysed by \citealt{Bharti13}) from the \ion{Ca}{II} H cube. These new arrays will be henceforth denoted `$10$TS' for ease. Each frame of $10$TS was then visually examined to identify bright elliptical features which were classed as potential SSUBs, with the co-ordinates of all candidates being recorded. Following this, binary maps were created for $2$\arcsec$\times2$\arcsec\ boxes surrounding each event with all pixels below an intensity of $\bar{x}+3\sigma$, where $\bar{x}$ and $\sigma$ denote the mean and standard-deviation of the background $10$TS FOV respectively, set to zero. Pixels above this value were set to one, mimicking the method of umbral micro-jet detection implemented by \citet{Bharti13}. Events which did not display this increase in intensity for two or more $10$TS frames were discarded. The remaining features, of which there were $54$ ($14$ in sub-region $1$ and $20$ in sub-regions $2$ and $3$), were considered to be the sample of SSUBs for this study. Again, we note that these SSUBs are likely to be analogous to the umbral micro-jets identified by \citet{Bharti13}. In Fig.~\ref{Position}, the FOV surrounding each of the sub-regions (white boxes) analysed are plotted for their first time-step. The over-laid white lines indicate spatial positioning of all identified SSUBs at their peak lengths.

\section{Results and discussion}
\label{Results}

\subsection{Properties of SSUBs in \ion{Ca}{II} H filtergrams}

Following their detection, the length of each event was measured at every third time-step in which it was visible. In the left frame of Fig.~\ref{Stats}, a histogram of peak lengths for all SSUBs is plotted. The mean length of $0.56$\arcsec\ ($\sigma$=$0.14$\arcsec), indicated by the dashed line, is around $0.1$\arcsec\ shorter than the mean of the peak umbral micro-jet lengths measured by \citet{Bharti13} ($0.67$\arcsec). It is likely that the slight discrepancy between the results of \citet{Bharti13} and those presented here is accounted for by the difference in $\mu$-angle and errors in the calculation of the mean length caused by small sample-sizes. These measurements support the results of \citet{Bharti13} that, in general, SSUBs are sub-arcsecond in size. Such extents are shorter than the typical $1$-$4$ Mm lengths of penumbral micro-jets (see, for example, \citealt{Katsukawa07}), the $1$ Mm lengths of short dynamic fibrils (\citealt{vanderVoort13}), and the $3$-$4$ Mm spans of the largest stable umbral fibrils (\citealt{Henriques13}).

In the central panel of Fig.~\ref{Stats}, a histogram of the lifetimes of the sample of events analysed in this article is plotted. Once again the mean for this parameter ($\bar{x}$=$44.2$ seconds; $\sigma$=$20$ seconds) is lower than measured for umbral micro-jets by \citet{Bharti13} ($\sim100$ s). We suggest that this difference is due to the improvement in the temporal resolution (by a factor of $16$) of the time-series presented here in comparison to the data studied by those authors. This is supported by the similarity of the shape of the distributions when the right-hand tail (the longer lived features) of the data presented by \citet{Bharti13} is not considered. The reason why no extended long-lifetime tail is found for the SSUBs analysed here is currently unclear but could be due to, for example, intrinsic differences in the studied sunspots (meaning different mean lifetimes could be expected) or human effects introduced during the collection of measurements. When compared to the $100$-$250$ second lifetimes of short dynamic fibrils (\citealt{vanderVoort13}), these SSUBs are shorter lived; however, they have comparable lifetimes to shock-driven umbral flashes (see, for example, \citealt{Beckers69,Socas00a}). Whether this is purely coincidental or offers some insights into the formation mechanism of SSUBs will be discussed later.

\begin{figure*}
\includegraphics[scale=0.55,trim={1.7cm 0 1.7cm 0}]{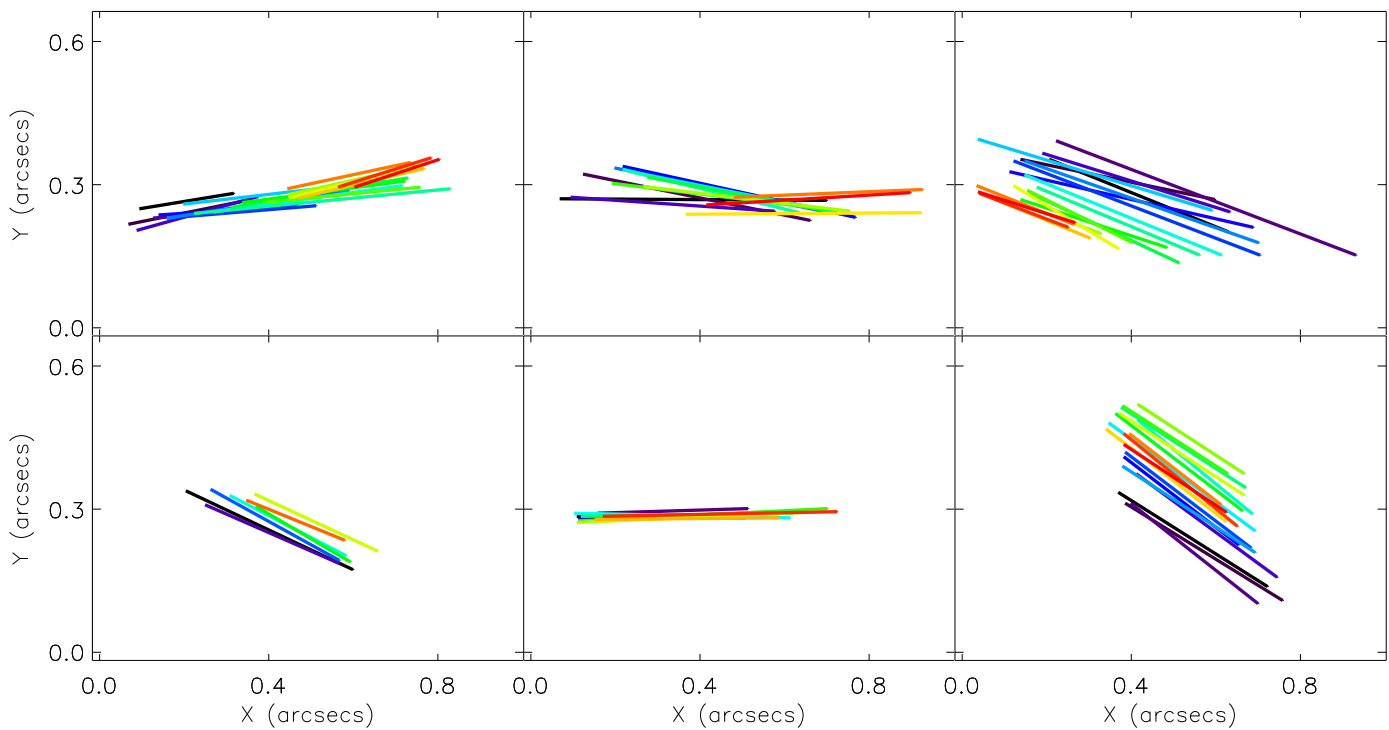}
\caption{Time evolution of six representative SSUB features. The evolution of colour, from dark purple (feature first identified) to green (middle of lifetime) to red (final frame in which feature is visible), shows the passage of time. The axis are plotted to give an indication of length (i.e. they are not solar co-ordinates and do not correspond to the axes of Fig.~\ref{Context}). This means that the initial positioning of the events in each frame is arbitrary. However, the relative shifts depict the measured motions of the events through time.}
\label{Extents}
\end{figure*}

A scatter plot of peak length against lifetime is included in the final panel of Fig.~\ref{Stats}. The dashed line indicates a linear fit calculated using the {\it linfit.pro} IDL procedure which returned a negligible gradient corresponding to approximately $0.0006$\arcsec\ of additional length per extra second of lifetime. The combination of this result with a calculated Pearson's correlation of just $0.09$ (computed using IDL's {\it correlate.pro}) provides strong evidence that peak length and lifetime have no statistical relationship within this SSUB sample. This lack of correlation is in contrast to the strong relationship between these variables found for short dynamic fibrils both outside and within sunspot atmospheres by, for example, \citet{dePontieu07} and \citet{vanderVoort13}. These results appear to challenge the legitimacy of the `jet' nomenclature for these events, justifying the need for the more generic term SSUB being used throughout this paper. Our results, therefore, support the assertions of both \citet{vanderVoort13} and \citet{Yurchyshyn14} that umbral micro-jets (which we call SSUBs in this article) and short dynamic fibrils (chromospheric spikes) are not the same phenomena.

In addition to lifetime and peak length, it is also of interest to measure the orientation of the SSUBs within our sample. \citet{Bharti13} suggested that the umbral micro-jets identified in their study were orientated parallel to the local penumbral micro-jets, meaning that their orientation was parallel to the local penumbral filaments. They used this as evidence of the jet-like nature of these features. As can be seen in Fig.~\ref{Position}, the orientation of SSUBs appears to differ between sub-region $3$ and the other analysed zones (agreeing with the results of \citealt{Bharti13}). Calculating the average orientations of SSUBs identified within each region yields $97.7^\circ$ ($\sigma$=$20^\circ$), $78.2^\circ$ ($\sigma$=$39^\circ$), and $-15.4^\circ$ ($\sigma$=$32^\circ$), measured anti-clockwise from the positive $x$-direction from Fig.~\ref{Context}, respectively. Statistically, the events in sub-region $3$ are orientated differently to features in the other two sub-regions ($t$-test $p$-values of $<0.05$). The SSUBs in sub-regions $1$ and $2$ are only significantly different if one considers a $p$-value below $0.1$. However, this more ambiguous statistical result could be expected due to the similarity in orientation of the local penumbral structures close to these sub-regions. Overall, we are able to assert that SSUBs in different regions of the umbra do indeed appear to have different orientations, which are comparable to the orientations of the local penumbral filaments.

To further understand these features, we also conducted an analysis of their evolution through time. Of the $54$ SSUBs discussed here, fewer than ten display any evidence of parabolic profiles when their length was plotted against time (discovered within short dynamic fibrils in sunspots by \citealt{vanderVoort13}). Considering events whose lengths did increase and decrease (i.e. potentially follow a parabolic path), it was rare for the foot-point to remain in one place. Instead, the event often progressed along an imaginery axis parallel to its orientation. In Fig.~\ref{Extents}, we plot the evolution of the spatial positioning of six typical SSUBs identified within these data, relative to an arbitrary centre point. The top left panel depicts one event whose length appears to be parabolic through time but which does not evolve from a single unmoving foot-point. Other panels display a sample of representative events which often appear to propagate away from their initial position as they evolve. Whether this spreading is comparable to that observed within umbral flashes (see, for example, \citealt{vanderVoort03}) will be discussed further later in this paper.

\begin{figure*}
\includegraphics[scale=0.52]{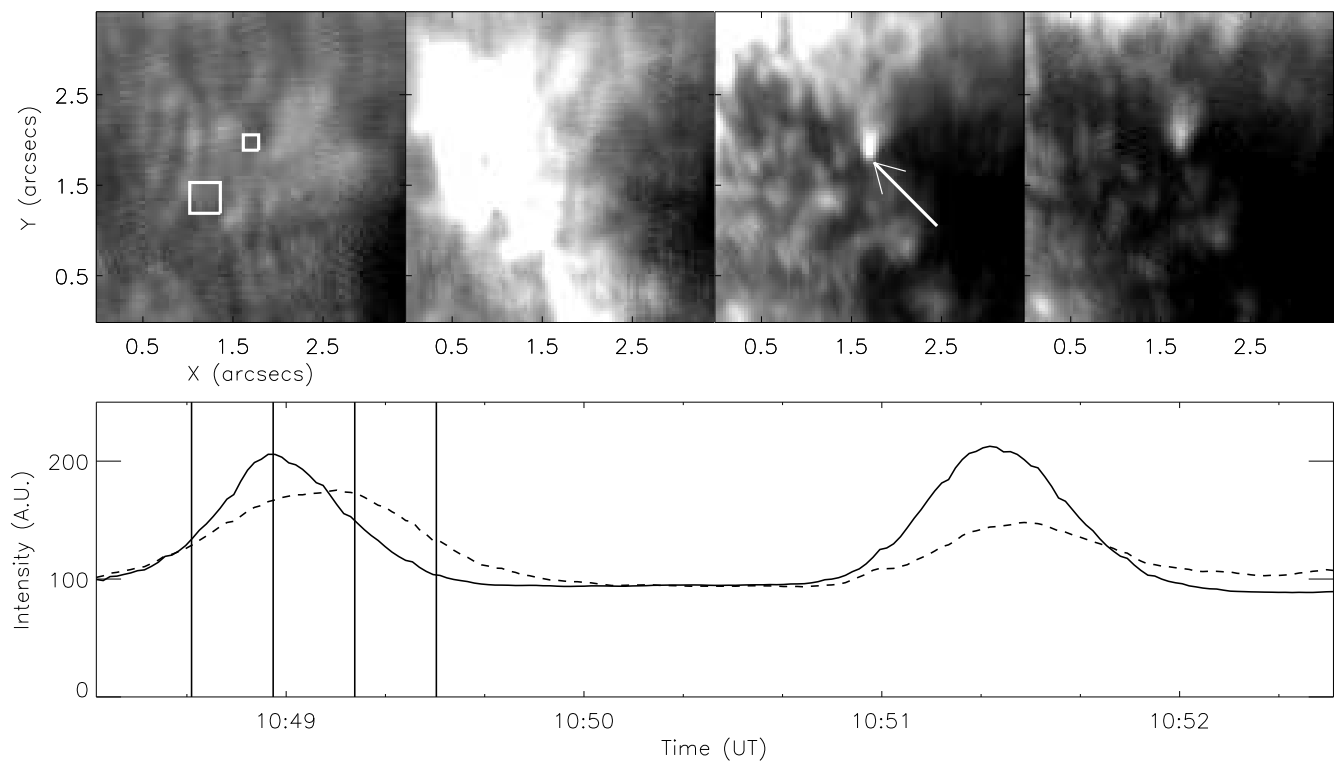}
\caption{(Top row) Evolution of a region of \ion{Ca}{II} H data displaying an umbral flash followed closely by a SSUB. Each frame is separated by $16.8$ seconds. The large and small boxes in the left-hand panel indicate the regions which are averaged to create the lightcurves plotted in the bottom panel. (Bottom row) A lightcurve plotting the intensity evolution of the umbral flash (solid line) and SSUB (dashed line). The solid vertical lines indicate the frames plotted in the top row. A second flash and SSUB pairing is also evident between $10$:$51$ UT and $10$:$52$ UT. A distinct time-lag between the peak intensities of the phenomena can be identified for both pairings.}
\label{Lightcurve}
\end{figure*}

Interestingly, the results of \citet{Bharti13} indicated that some umbral micro-jets occurred co-spatially with umbral flashes. These authors did, however, state that other umbral micro-jets were clearly observed in quiet regions of the umbra and suggested that this was evidence of a difference between these events and the fine structuring of umbral flashes observed by, for example, \citet{Socas09}. Within our sample, umbral flashes were observed to form co-spatial and co-temporal with $28$ out of the $54$ SSUBs. For the purposes of this study, co-temporal was defined as within one minute of the on-set or disappearance of the SSUB. The top row of Fig.~\ref{Lightcurve} plots four panels depicting the evolution of a flash and SSUB pairing (with each frame separated by $16.8$ seconds). In the second panel the flash occurs, which disappears by the third panel where the SSUB is clearly observed (indicated by the white arrow). The fourth panel depicts a later frame, after the SSUB had decreased in intensity and spread to cover a larger area. Large and the small white boxes in the first panel indicate the regions selected to construct a lightcurve for the umbral flash and SSUB, respectively.

The constructed lightcurve is displayed in the bottom panel of Fig.~\ref{Lightcurve}, with the vertical black lines indicating the time-steps plotted in the top row. This plot clearly depicts a lag between the peak intensity of the flash (solid line) and the SSUB (dashed line) of around $10$-$20$ seconds. However, within the sample of $28$ SSUBs which occur co-spatially with flashes, both positive and negative lags are observed, indicating that no preferred ordering exists for whether a flash or SSUB occurs first. It is interesting that a second flash and SSUB pairing is also evident within this lightcurve at approximately $10$:$51$:$30$ UT. This repetitive behaviour is not unique to this example, but was not common within our sample. What leads to the formation of multiple SSUBs at the same location over a short period of time will be explored in the remainder of this article. It is likely that this behaviour is indicative of a sustained inhomogeneity in the umbra, or due to a repetition of the driving mechanism responsible for this increased intensity.

\begin{table*}
\caption{Number of SSUBs observed co-spatial to umbral flashes and umbral dots in the three sub-regions analysed here.}
\begin{center}
\begin{tabular}{p{3cm} p{2cm} p{2cm} p{2cm} p{2cm} p{2cm} p{2cm}}
 \hline
&  \multicolumn{2}{c}{Sub-Region 1} & \multicolumn{2}{c}{Sub-Region 2} & \multicolumn{2}{c}{Sub-Region 3} \\
 \hline
 & Umbral Flash & non-Umbral Flash & Umbral Flash & non-Umbral Flash & Umbral Flash & non-Umbral Flash \\
 \hline
Umbral Dot & 4 & 3 & 8 & 3 & 6 & 8 \\
Non Umbral Dot & 2 & 5 & 6 & 3 & 2 & 4 \\
 \hline
\end{tabular}
\end{center}
\label{Stats_table}
\end{table*}

\subsection{CRISP data - Spectral and polarimetric signatures}

To further our analysis we now consider the data collected in the red beam. \citet{Bharti13} discussed a potential relationship between the spatial formation of umbral micro-jets and the occurrence of umbral dots in the photosphere. Therefore, we focus our initial analysis on the wide-band data collected in tandem with the SST/CRISP spectro-polarimetric scans to infer whether any links can be deduced between the SSUBs identified in this study and photospheric umbral dots. Overall, $32$ of the $54$ ($59$ \%) identified SSUBs appeared to be orientated towards local umbral dots with apparent distances of $<2\arcsec$. This value is close to the $56$ \% noted by \citet{Bharti13} for umbral micro-jets. Of the $32$ events potentially linked to umbral dots, no differences in frequency were found between SSUBs co-spatial to umbral flashes and those occurring in quieter regions with $18$ (out of $28$) and $14$ (out of $26$) potential links observed, respectively. There also appeared to be no difference in the frequency of this potential spatial relationship between the sub-regions analysed, with around $50$-$75$ \% of SSUBs ($7$, $11$, and $14$ in regions $1$, $2$, and $3$, respectively) occurring close to umbral dots across the sunspot.

\begin{figure*}
\includegraphics[scale=0.51]{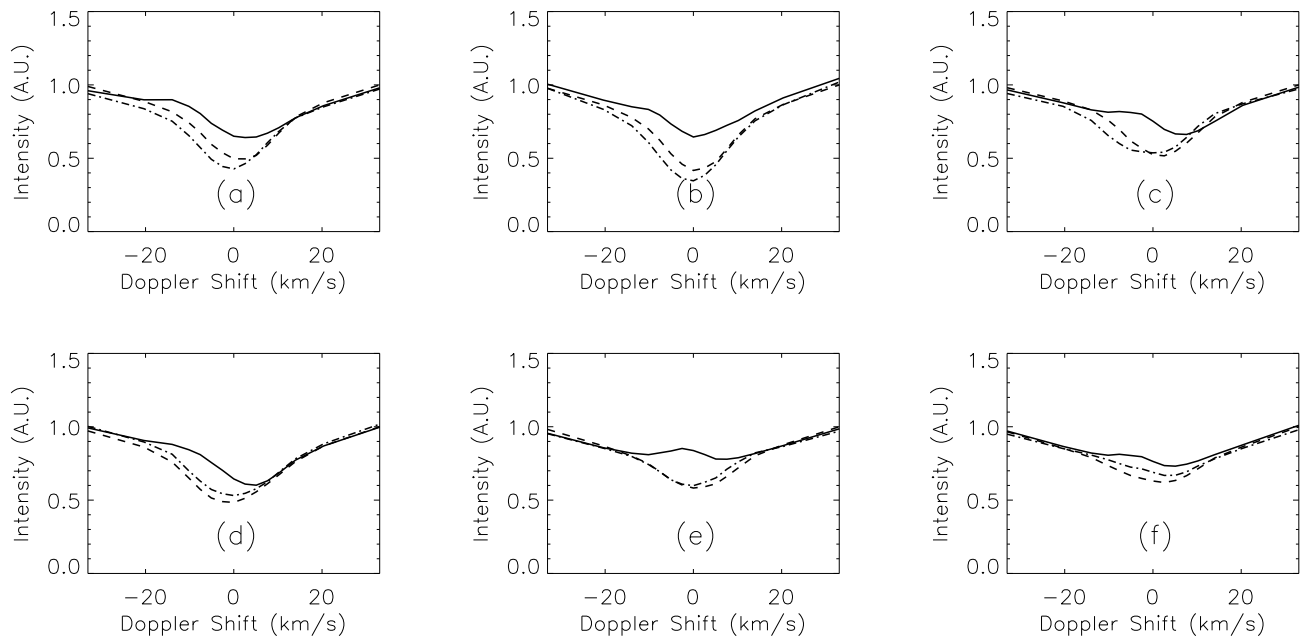}
\caption{(a-c) \ion{Ca}{II} $8542$ \AA\ line profiles (averaged over 5 pixels and with Doppler shift referenced against the \ion{Ca}{II} $8542$ \AA\ line core) of three SSUBs which occurred close to an umbral flash. (d-f) The same as the top row but for three SSUBs not visibly associated with an umbral flash. For each frame, the solid line plots the profile during the occurrence of the SSUB, the dashed line plots the scan one minute prior to the formation of the feature, and the dot-dashed line corresponds to the scan measured one minute after the SSUB.}
\label{Profiles}
\end{figure*}

Of the $22$ events not occurring near obvious umbral dot structuring, ten belonged to the sample containing those SSUBs co-temporal to a local umbral flash. The remaining $12$ SSUBs ($5$, $3$, and $4$ in sub-regions $1$, $2$, and $3$, respectively) were neither associated with an umbral flash nor an umbral dot. Our results do not, therefore, indicate that SSUBs form preferentially over umbral dots. In addition, we find no evidence that those SSUBs appearing to occur co-temporally with umbral flashes are more or less likely to coincide with umbral dots nor is there clear evidence that two populations of SSUBs exist (e.g. SSUBs which form as small-scale flash-like events and upward flow-generated SSUBs). We note, however, that these results do not provide conclusive evidence as to whether a relationship exists between SSUBs and umbral dots. Associating chromospheric events with a photospheric source is difficult, especially on scales as small as for the events studied here. It is possible, for example, that umbral dot structuring occurs on scales below those sampled by these data, or that some SSUBs are linked to umbral dots which are situated more than $2$\arcsec\ away. A larger statistical sample than the combined $77$ SSUBs (inclusive of those features analysed by \citealt{Bharti13} plus the SSUBs presented here) could provide further insights to this possible relationship.

In addition to this, $35$ of the SSUBs analysed here occured co-spatial to a compact brightening in the \ion{Ca}{II} $8542$ \AA\ line profile, often strongest in the near blue wing or line core. Due to potential problems caused by the lower cadence of these SST/CRISP observations compared to the blue data, the dynamic nature of umbral chromospheres (e.g. the rapid intensity variations caused by umbral flashes), and the short lifetimes of these features, we now focus on a sample of six representative events which have signatures in both the blue and red data. Of these, three belong to those SSUBS apparently linked to umbral flashes and three are not directly observed co-spatial to a flash. These six events could offer some insights into the spectral signatures of all SSUBs in the \ion{Ca}{II} $8542$ \AA\ line. However, it is possible that the visibility of these specific features in the red beam in itself could be an indicator that this smaller sample are different to those $19$ SSUBs which show no signature in \ion{Ca}{II} $8542$ \AA\ line scans. 

\begin{figure*}
\includegraphics[scale=0.51]{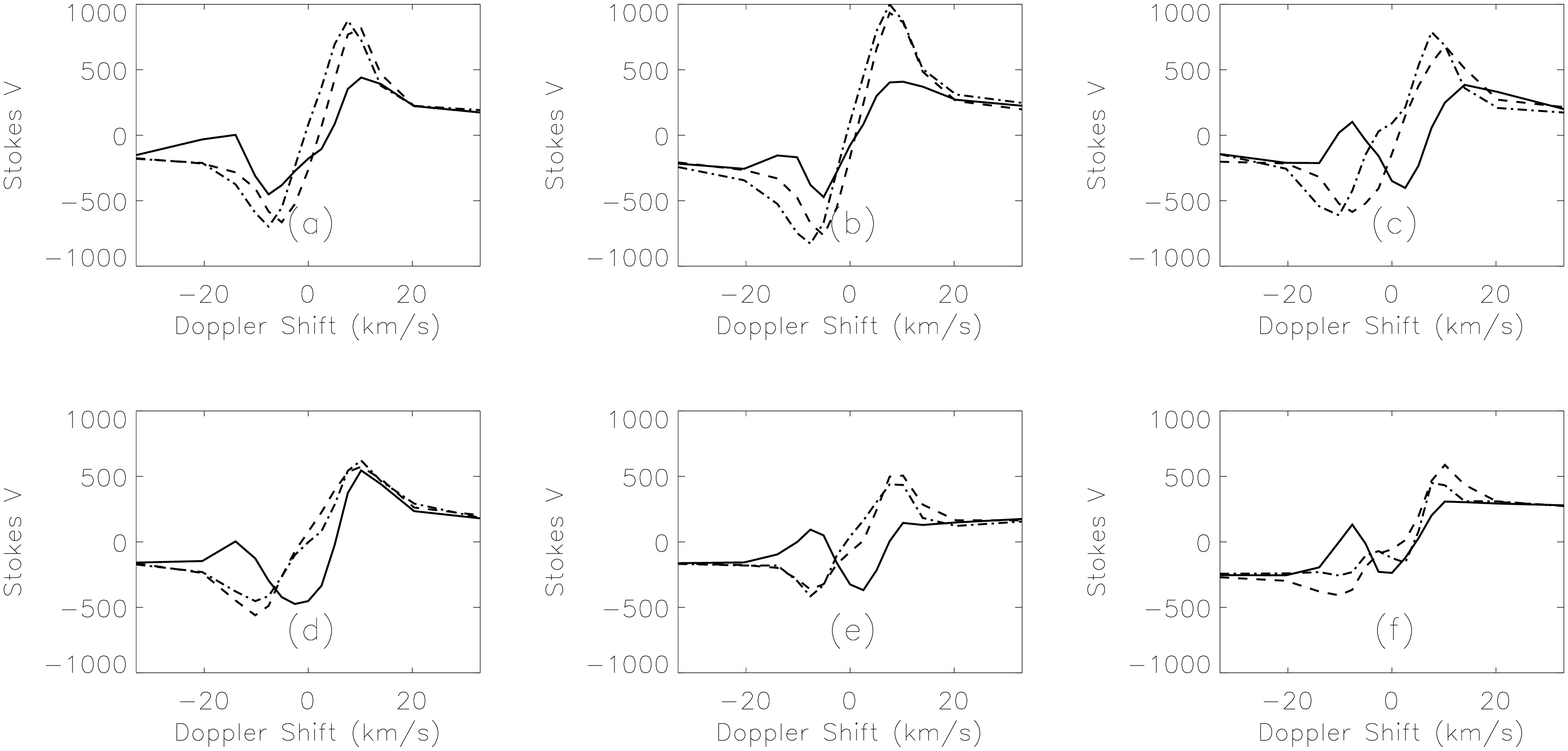}
\caption{Same as Fig.~\ref{Profiles} but for the Stokes V component.}
\label{Profilesv}
\end{figure*}

In Fig.~\ref{Profiles} we plot normalised profiles sampled across the \ion{Ca}{II} $8542$ \AA\ line for the six features selected for analysis. The solid lines in each frame show the line profile created from the scan co-temporal to the occurrence of the SSUB, the dashed line is the scan one minute prior to the SSUB, and the dot-dashed line the scan one minute following the SSUB. In the top row we plot those SSUBs which appear close to umbral flashes, (both spatially and temporally), and the bottom row depicts profiles of SSUBs which appear isolated in the umbra. Fig.~\ref{Profiles}(a) corresponds to the SSUB presented in Fig.~\ref{Lightcurve}. It is immediately evident that increases in intensity in the \ion{Ca}{II} $8542$ \AA\ line core can be identified in all six examples (a-f). Indeed, all of these plots include an asymmetric bump in intensity in the near blue wing, apparently similar to the umbral flash profiles discussed by, for example, \citet{delaCruzRodriguez13}. We suggest this is evidence that SSUBs are in fact increases in intensity caused by the formation of shocks in the lower solar atmosphere. The slight differences in the shapes of the profiles could, therefore, be caused by the progression of the intensity increase from the blue to the red component of the line, as was first seen by \citet{Beckers69}.

The \ion{Ca}{II} $8542$ \AA\ line profiles presented here support the assertions made in the previous section that SSUBs, detected in a similar manner to the umbral micro-jets discussed by \citet{Bharti13}, do not appear to be jet-like in nature. Of course,  one should be wary of the summation of profiles within one spatial pixel. It is possible, for example, that the shock profile observed co-spatial to these SSUBs is, in fact, caused by a different, unrelated event (such as an umbral flash or running penumbral wave) occurring co-spatially with the SSUB. We suggest this is unlikely for the bottom row, however, due to the lack of obvious flash signal co-spatial to these events. In addition to this, the \ion{Ca}{II} $8542$ \AA\ profiles do not display multiple components which (if they existed) would indicate complex stratification in the local atmosphere (\citealt{Toriumi17}) and are resolved over multiple pixels implying that the spatial extent of the shock is comparable to the sizes of the SSUBs themselves. The current evidence is compelling in its support of the nature of SSUBs as signatures of shocked material.

In addition to the Stokes I component, both \citet{Socas00a} and \citet{delaCruzRodriguez13} discussed the influence of the umbral flashes on the Stokes V profile. Therefore, we proceed in a similar manner by analysing the measured Stokes V components of SSUBs. In Fig.~\ref{Profilesv} we plot the Stokes V profiles corresponding to the respective events studied in Fig.~\ref{Profiles}. Each of these frames depicts a major change in the shape of the profile during the SSUB, with all events clearly displaying the polarity reversal prevalent within umbral flashes to different degrees. The similarity of these plots to the Stokes V profiles plotted in Fig.~4 of \citet{delaCruzRodriguez13} is striking, further supporting the shock hypothesis for the formation mechanism of SSUBs. 

\begin{figure}
\includegraphics[scale=0.56]{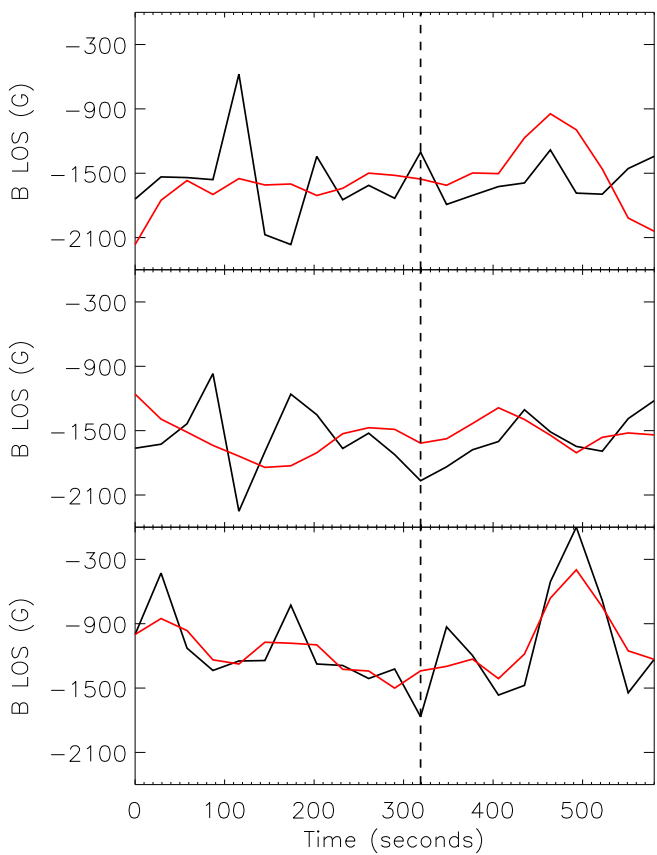}
\caption{Estimates of the line-of-sight magnetic field strength as a function of time calculated using the weak-field approximation at the positions of three SSUBs (specifically a, c, and d from Fig.~\ref{Profiles}). The black line plots the frame-by-frame calculation and the red line the smoothed approximation over three frames. The dashed line indicates the time at which the solid profiles within Figs.~\ref{Profiles} and  \ref{Profilesv} were  sampled.}
\label{WFA}
\end{figure}

The physical perturbation responsible for this reversal was hypothesised to be thermodynamic rather than magnetic by \citet{delaCruzRodriguez13}. We, therefore, use the weak-field approximation (see, for example, \citealt{Stenflo13}, \citealt{delaCruzRodriguez13}) to calculate the line-of-sight magnetic field strength as a function of time for a sample of SSUBs to look for large-scale changes to the magnetic field during these events. In Fig.~\ref{WFA}, we plot the approximated magnetic field strength for three of the six SSUBs shown in Figs.~\ref{Profiles} and \ref{Profilesv} (specifically features a, c, and d). The black line plots the estimated field strength for each time-step, calculated by fitting the gradient of the Stokes I component to the Stokes V component, with the line-of-sight magnetic field as the variable. The red lines show the magnetic field smoothed over three frames, while the dashed line indicates the time of the occurrence of the SSUB. As with the flash profiles analysed by \citet{delaCruzRodriguez13}, the variations in line-of-sight magnetic field occurring co-temporal to the SSUB do not show one specific trend, indicating that thermodynamic effects (e.g. changes in temperature or density) are the cause of the perturbations to the Stokes profiles during the SSUBs.

To complete the spectral analysis of these events, we also examined the H$\alpha$ line core images co-spatial and co-temporal to these SSUBs. No noticeable signature (e.g. compact brightening or dark region) was observed in the H$\alpha$ line core for any of the features examined in our sample. It is possible that this lack of signature implies that the shock which lead to the appearance of the SSUB in the \ion{Ca}{II} H and \ion{Ca}{II} $8542$ \AA\ lines forms in the lower chromosphere or even in the upper photosphere, below the formation height of the H$\alpha$ canopy, however, this is currently only speculation. We shall examine this hypothesis in future work using H$\alpha$ line scans and inversions of the atmosphere calculated using the NICOLE code (see, for example, \citealt{Socas15}).

\subsection{Links between SSUBs and short dynamic fibrils}

The final analysis presented here regards a link between these SSUBs and short dynamic fibrils. Short dynamic fibrils are thought to be formed by the upward propagation of corrugated shock fronts caused by varying thermodynamic properties in the horizontal plane (see, for example, \citealt{Heggland11}), potentially on scales comparable to the SSUBs analysed here.  In Fig.~\ref{Fibrils} we depict the evolution of the FOV surrounding two SSUBs (including the event plotted in Fig.~\ref{Lightcurve}) for six time-steps. Each row plots one of three different line positions (indicated on the first panel of each respective row) and each column is separated by $29$ seconds. The SSUBs \ion{Ca}{II} $8542$ \AA\ response can clearly be seen in the fourth and sixth columns, respectively, in both the $-0.217$ \AA\ and line core images. White arrows indicate the co-spatial short dynamic fibrils during their progression from the blue to the red wing. Both SSUBs are observed at the foot-points of the sunspot fibrils during the red-shifted portion of their evolution. This result potentially implies a link between the down-flowing fibrilar material and the formation of the SSUB,or a modulation of an already present shock-wave at these locations due to the localised thermodynamic conditions. The lifetimes ($\sim180$ seconds) and lengths ($\approx400$ km) of these events are comparable to both short dynamic fibrils and chromospheric spikes previously discussed in the literature (\citealt{vanderVoort13} and \citealt{Yurchyshyn14}, respectively). 

Overall, $17$ of the $35$ SSUBs with clear signatures in the \ion{Ca}{II} $8542$ \AA\ line profile are observed to form at the foot-point of dynamic fibril events. We note that this estimate is likely only a lower limit due to the obscuration of some events, for example, by bad seeing or the occurrence of umbral flashes. The appearance of the sunspot fibrils prior to the SSUBs in these data, however, is intriguing. We, therefore, outline two hypotheses that could account for their appearance. One potential explanation is that denser material at the location of the dynamic fibril delays the formation of the shock, leading to a lag in the detection of the flash at these locations, with the delayed flash then being identified as a SSUB (as shown in Fig.~\ref{Lightcurve}). A second explanation is that these SSUBs are the signatures of shocking in the lower solar atmosphere caused by the return of fibrilar material. Support for this scenario includes the results that SSUBs generally occur co-temporally with the downward (red-shifted) portion of the lifetime of co-spatial sunspot fibrils and are, therefore, unlikely to be a direct result of any shock which excites the fibril initially, as well as the lack of correlation between SSUB formation and the occurrence of umbral flashes, as was discussed in the previous sections. This scenario could also explain the periodic repetition of the SSUBs identified in Fig.~\ref{Lightcurve} due to the periodic repetition of short dynamic fibrils (\citealt{vanderVoort13}). Recent inversion work has found signatures of down-flowing material co-spatial to umbral flashes and to dark umbral fibrils. Furthermore, \ion{Ca}{II} $8542$ \AA\ flash-like profiles, the same observed for SSUBs, were reproduced with downflowing atmospheres using NLTE radiative transfer, which is consistent with the scenario of flash-like profiles co-spatial to down-falling short dynamic fibril material (Henriques et al., 2017, Submitted). It is possible that both of these hypotheses are each responsible for a sub-set of SSUBs. We plan to test both of these hypotheses in future research.

\section{Conclusions}
\label{Conclusions}

\begin{figure*}
\includegraphics[scale=0.51]{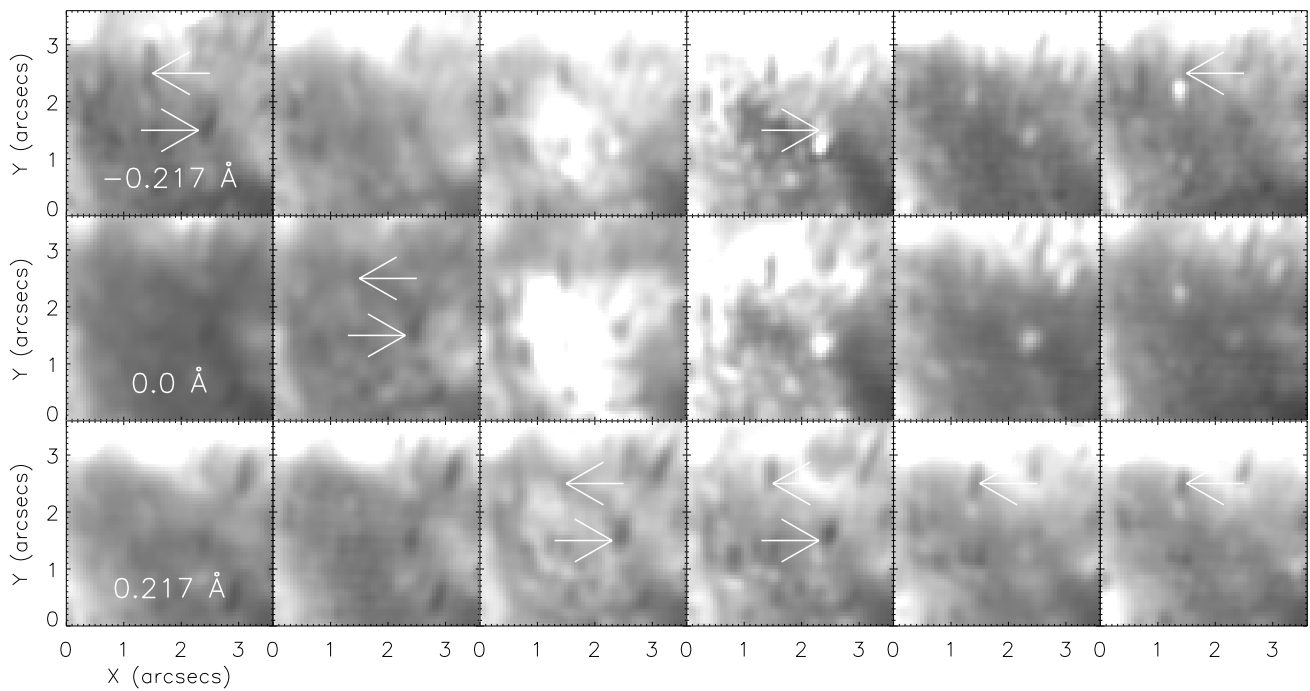}
\caption{(Top row) The evolution of a FOV around two SSUBs (including the event plotted in the top left panel of Fig.~\ref{Profiles}) at $-0.217$ \AA\ into the blue wing of the \ion{Ca}{II} $8542$ \AA\ line. Each column is separated in time by about $29$ seconds. (Middle row) Same as the top row but for the \ion{Ca}{II} $8542$ \AA\ line core. (Bottom row) Same as above but for $0.217$ \AA\ into the red wing of the \ion{Ca}{II} $8452$ \AA\ profile. Arrows indicate the positions of the co-spatial short dynamic fibrils. The SSUBs are clearly visible in the fourth and sixth columns of the $-0.217$ \AA\ and line core images (indicated by white arrows in the fourth and sixth frames of the top row).}
\label{Fibrils}
\end{figure*}

A wide variety of features have been discovered within sunspot atmospheres in recent years, ranging from partially understood short dynamic fibrils (\citealt{vanderVoort13}) to the intriguing umbral micro-jets (\citealt{Bharti13}). How this new set of features fit into the complex zoo of events, such as umbral flashes and umbral dots, already observed within sunspot umbrae is still unknown. We have clearly shown that SSUBs can be observed within \ion{Ca}{II} H line core filtergrams as well as (for a sub-set) in \ion{Ca}{II} $8542$ \AA\ line profiles in the umbra of a large sunspot. Furthermore, evidence is provided that these features form as a response of the atmosphere to shocks, leading to similar spectral signatures as umbral flashes. It is likely that these SSUBs exist due to localised structuring (similar to the fine-scale organisation within umbral flashes observed by \citealt{Socas09}) but are formed, not due to a delay of the umbral flash shock-front, but by down-falling material from short dynamic sunspot fibrils. Below we briefly reiterate our key findings.

Initially, our work focussed on discerning a sample of SSUBs within \ion{Ca}{II} H filtergrams and understanding their basic properties. A total of $54$ events were identified (the spatial positioning of which can be found in Fig.~\ref{Position}), which appeared to be similar in nature to the umbral micro-jets discovered by \citet{Bharti13}. The SSUBs analysed here had average values for both peak length and lifetime of $0.56$\arcsec\ ($\sigma$=$0.14$\arcsec) and $44.2$ seconds ($\sigma$=$20$ seconds), respectively, as can be seen in the first two panels of Fig.~\ref{Stats}. The orientation of the SSUBs was found to be specific to the region of the sunspot in which they were observed, indicating that these events could be influenced by the orientation of the local magnetic field. Again, this result is in agreement with the umbral micro-jets studied by \citet{Bharti13}. It is therefore highly likely that the SSUBs analysed here correspond to the umbral micro-jets discussed by \citet{Bharti13}. Interestingly, no correlation was found between peak length and lifetime for the SSUBs, challenging the `jet' nomenclature originally applied to these events. This result is in support of the assertions of \citet{vanderVoort13} and \citet{Yurchyshyn14}, who stated that the short dynamic fibrils and chromospheric spikes observed in sunspots were dissimilar to the umbral micro-jet phenomena.

Analysis of the morphological evolution of the SSUBs with time showed little evidence of either parabolic or ballistic profiles, with most features appearing to brighten along their whole lengths almost immediately (as can be seen in Fig.~\ref{Extents}). This property is similar to the rapid occurrence of umbral flashes (theorised to form due to shocks: \citealt{Bard10}), where a large region brightens from one frame to the next. In addition it was found that some SSUBs appeared to recur in the same spatial location (see Fig.~\ref{Lightcurve}) with a periodicity close to that found for the local umbral flashes. We suggest, therefore, the possibility that the umbral micro-jets discussed by \citet{Bharti13} are not actually jet-like in nature, but are instead shock-driven. 

To further test the possibility that SSUBs could be shock related, we also analysed their signature in the \ion{Ca}{II} $8542$ \AA\ line profile. Overall, $35$ SSUBs were identified to have compact regions of increased intensity close to the core of this line. The line profiles of six of these SSUBs were then analysed in detail (plotted in Fig.~\ref{Profiles}), appearing to depict umbral flash-like profiles (similar to those presented by \citealt{delaCruzRodriguez13}), with increases in intensity and asymmetry about the line core. Only one event (Fig.\ref{Profiles}e) did not display strong evidence of a blueshift around its core. However, this could be due to the relatively sparse temporal sampling of the profile in comparison to the lifetime of the feature, meaning the shock profile had become more redshifted (as discussed by \citealt{Beckers69}). The lack of any co-spatial signature in the H$\alpha$ line core could suggest a formation height of the shocks leading to this increased intensity at around the top of the photosphere. However, this is currently only speculation and will be investigated in further work.

Reversals in the Stokes V component during the occurrence of the SSUB were also observed (see Fig.~\ref{Profilesv}), providing further evidence for the shock-driven formation mechanism. Recently, \citet{delaCruzRodriguez13} used the weak-field approximation to infer changes to the line-of-sight magnetic field over time. We conducted a similar analysis on data co-spatial to the SSUBs, finding little change in the magnetic field strength during the visibility of the events (as was shown in Fig.~\ref{WFA}). This suggests that thermodynamic changes in the local plasma are the main driver of the increased emission and flash-like profiles co-spatial to SSUBs, similar to the results obtained for umbral flashes by \citet{delaCruzRodriguez13}.

Finally, $17$ of the SSUBs observable in the \ion{Ca}{II} $8542$ \AA\ line profile appeared to be co-spatial with the foot-point of short dynamic fibrils, such as those discussed by \citet{vanderVoort13}. Short dynamic fibrils were identified as dark extensions in \ion{Ca}{II} $8542$ \AA\ data, transitioning from the blue to the red wing during their lifetime. These transitions were clearly evident from our imaging data (see Fig.~\ref{Fibrils}), with the SSUBs usually occurring whilst the fibril was apparent in the red wing of the line. We suggest that this co-spatial relationship could be evidence of shock formation in the lower portion of the fibril structure caused by the return of fibrilar material to the lower atmosphere due to gravity or due to a modulation of an already present shock by the localised thermodynamic properties at these locations caused by the presence of the fibril. We will test this hypotheses in follow-up work.

\begin{acknowledgements}
We thank the UK Science and Technology Facilities Council (STFC) for the support received to conduct this research. This research was supported by the SOLARNET project (\url{http://www.solarnet-east.eu}). The Swedish $1$-m Solar Telescope is operated on the island of La Palma by the Institute for Solar Physics of Stockholm University in the Spanish Observatorio del Roque de los Muchachos of the Instituto de Astrofísica de Canarias. We would like to thank the anonymous reviewer for comments which improved this manuscript.
\end{acknowledgements}

\bibliographystyle{aa}
\bibliography{SunspotFibrils}

\end{document}